\documentstyle[epsf,graphics,a4,12pt]{article}
\begin{document}\baselineskip=18pt
\def\be{\begin{equation}}
\def\ee{\end{equation}}
\def\bearst{\begin{eqnarray*}}
\def\eearst{\end{eqnarray*}}
\def\peleven{\parbox{11cm}}
\def\peffec{\peight{\bearst\eearst}\hfill\peleven}
\def\pspace{\peight{\bearst\eearst}\hfill}
\def\ptwelve{\parbox{12cm}}
\def\peight{\parbox{8mm}}
\def\bear{\begin{eqnarray}}
\def\eear{\end{eqnarray}}
\def\E{{\rm e}}
\input epsf.tex
\newcommand{\slp}{\raise.15ex\hbox{$/$}\kern-.57em\hbox{$\partial$}}
\newcommand{\sla}{\raise.15ex\hbox{$/$}\kern-.57em\hbox{$a$}}
\newcommand{\slA}{\raise.15ex\hbox{$/$}\kern-.57em\hbox{$A$}}
\newcommand{\slB}{\raise.15ex\hbox{$/$}\kern-.57em\hbox{$B$}}
\newcommand{\slD}{\raise.15ex\hbox{$/$}\kern-.57em\hbox{$D$}}
\newcommand{\slb}{\raise.15ex\hbox{$/$}\kern-.57em\hbox{$b$}}
\newcommand{\slW}{\raise.15ex\hbox{$/$}\kern-.57em\hbox{$W$}}
\font\grrm=cmbx10 scaled 1200
\font\vb=cmbx10 scaled 1440
\font\bigcal=cmsy10 scaled 1200
\def\eightpoint{\def\rm{\fam0\eightrm}}
\def\flex{\raise 6pt\hbox{$\leftrightarrow $}\! \! \! \! \! \! }
\def\tr{ \mathop{\rm tr}}
\def\atanh{\mathop{\rm atanh}}
\def\Tr{\mathop{\rm Tr}}
\def\dal{\Box} 
\def\Natural{\hbox{\hskip 1.5pt\hbox to 0pt{\hskip -2pt I\hss}N}}
\def\Integer{\>\hbox{{\sf Z}} \hskip -0.82em \hbox{{\sf Z}}\,}
\def\Rational{\hbox{\hbox to 0pt{\hskip 2.7pt \vrule height 6.5pt
                                  depth -0.2pt width 0.8pt \hss}Q}}
\def\Real{\hbox{\hskip 1.5pt\hbox to 0pt{\hskip -2pt I\hss}R}}
\def\Complex{\hbox{\hbox to 0pt{\hskip 2.7pt \vrule height 6.5pt
                                  depth -0.2pt width 0.8pt \hss}C}}
\def \ln {{\rm ln}\, }
\def \cotg {\rm cotg }
\vskip 1cm
\begin{tabbing}
\hskip 11.5 cm \= \\
\>hep-th/9808003\\
\>August 1998
\end{tabbing}
\vskip 1cm
\begin{center}
{\Large\bf Screening in three-dimensional QED with arbitrary fermion mass}
\vskip 1.2cm
{\large \bf E. Abdalla$^{a,1}$, 
R. Banerjee$^{b,2}$ and C. Molina$^{a,3}$\\
\vskip 0.4cm
{\it $^a$Instituto de F\'\i sica-USP, C.P.66318, 05315-970
S. Paulo, Brazil}\\
{\it $^b$S.N. Bose National Centre for Basic Sciences
Block JD, Sector III, Salt Lake City, Calcutta 700.091 India}\\
\sl $^1$ eabdalla@fma.if.usp.br\\
$^2$ rabin@boson.bose.res.in\\
$^3$cmolina@fma.if.usp.br}
\end{center}
\abstract

We compute the quark--antiquark potential in three dimensional
massive Quantum Electrodynamics for arbitrary  fermion mass. 
The result indicates that screening
prevails for any quark masses, contrary to the classical expectations,
generalizing our previous result obtained for large masses. We also
test the validity of several approximation schemes using a detailed 
numerical analysis. The classical result is still reproduced for 
small separation of the quarks.

\vfill\eject

\section{Introduction}

A proper study of the problem of screening  and confinement is of
considerable importance in our understanding of gauge theories.
To avoid the complexities of four dimensions these studies are usually 
confined to lower dimensions. 
In this framework, a deep physical interpretation has been achieved. Indeed, 
in two--dimensional QED\cite{klaus-heinz}, one obtains screening in the 
massless case, but confinement in the massive quark case, realizing the 
expected picture. 

For QCD in two dimensions Gross et al \cite{gross} were the first to discuss
the subject. If dynamical fermions and test charges are in different
representations, they find screening or confinement in some particular cases
depending on whether the fermion is massless or massive.
A similar conclusion in an identical setting has been
arrived at for the massless case in \cite{frso}. If, on the other hand, all
fermions are in the fundamental representation, then screening
prevails independently of the quark mass \cite{amz}.

General inquires in two dimensional gauge theories have been performed
recently by several authors \cite{semenov}, concerning the $\theta$ 
vacuum structure, screening, confinement and chiral condensates. In 
three dimensions  related questions were studied in \cite{semqed3}.

In three dimensional space-time, for an abelian gauge group, the question
of screening versus confinement has been recently analysed for large fermion 
masses\cite{abdban} in which case the fermionic determinant can be computed 
as a series in the inverse mass. The conclusion was that, contrary to
classical expectations, the theory is in the screening phase. Although this
is expected from the fact that a Chern-Simons  term develops and there
is a topological mass generation, it is a further indication that the
dynamics of gauge fields and the deep problem of screening versus
confinement is far from being settled by a simple inspection of the
classical behaviour of the theory. In the case of three dimensional
QED, the outcome reveals that the vacuum polarization is once more
capable of developing configurations that screen the external quarks,
presumably modifying the dynamics of quark-antiquark bound-states.

Here we extend the analysis of our previous work \cite{abdban} in order
to include all values of the fermion mass parameter. An explicit 
expression for the quark-antiquark potential is obtained following the usual
ideas of bosonisation\cite{klaus-heinz,coleman-mandelstam} but an analytic 
form cannot be obtained, and we resort to the use of numerical analysis. The 
results show that the screening phase obtained in the large mass 
limit\cite{abdban} persists for any value of the mass parameter (including 
vanishing mass). Next, the validity of certain approximation 
schemes\cite{fosco} is tested. Using these approximations a simple form of 
the quark-antiquark potential can be given, which is compared with the exact 
form. We still obtain screening; moreover the behaviour of the functions
is very accurately described by the proposed approximations.

The paper is divided as follows. In section two we present the computation
of the potential and the numerical results. We draw the
potential for different values of the mass parameter, showing that
its form is essentially the same in the whole range of real values for the
mass term, leading to the screening phase. Furthermore, we test the
approximation forms of all the functions necessary for computing the potential.
Section three is reserved for conclusions and discussions.

\section{Computation of the quark-antiquark potential}

The partition function of three dimensional massive QED
in the covariant gauge, in the presence of an external source $J^\mu$, 
is given by
\begin{eqnarray}
Z &= &\int d[\psi, \bar{\psi},
A_{\mu}]\delta(\partial_{\mu} A^{\mu}){\em exp} \left\{\
i\int d^3 x [\bar{\psi}(i\partial\!\!\!/\, - m -  e
A\!\!\!\!/\,)\psi \nonumber \right.\\ 
& & \left.- \frac{1}{4} F^{2}_{\mu\nu} + A_\mu J^\mu ]\ \right\}
\label{partition-function}
\end{eqnarray}
where $F_{\mu\nu}$ is the field tensor, $F_{\mu\nu}=\partial_\mu A_\nu -
\partial_\nu A_\mu $.

The bosonised version of the above defined action in the weak coupling
approximation is given by the expression\cite{boso3,bm}
\begin{eqnarray}
Z &= &\int dA_{\mu}\delta(\partial_{\mu}A^{\mu})exp\  \left\{ i
\int d^{3}x \times\right.%
\nonumber\\
&&%
\left.\lbrack \frac 12
A_{\mu}\Pi^{\mu\nu}A_{ \nu} -\frac 14 F_{\mu\nu}^2
+ A_\mu J^\mu  +.....\rbrack\right\} \quad ,\label{bosoaction}
\end{eqnarray}
where the dots stand for non quadratic terms in the gauge field
$A_\mu$. This result will describe the partition function of
the Maxwell-Chern-Simons\cite{djt} theory in the covariant gauge
in the infinite mass limit\cite{abdban}, as we see from the explicit
expression for the self-energy of the gauge field, $\Pi_{\mu\nu}$, given
by the expression
\be
\Pi_{\mu\nu}= H(p)i\epsilon_{\mu\nu\rho}\frac {p^\rho}{p^2}
+\left[G(p) + p^2\right]
\left(g_{\mu\nu}-\frac {p_\mu p_\nu}{p^2}\right)
\quad ,\label{self-gauge}
\ee
where the functions $G$ and $H$ are given by the forms
\bear
H(p)&=& - \frac {e^2p^2}{4\pi}
\int_0^1dt \frac m{\lbrack m^2 - t(1-t)p^2\rbrack^{1/2}}
\quad ,\label{h-func}\\
G(p)&=& -p^2 -\frac { e^2 p^2}{2\pi}
\int_0^1dt\, \frac {t(1-t)}{\lbrack m^2 - t(1-t)p^2\rbrack^{1/2}}
\quad , \label{g-func}
\eear
and $p=\sqrt{-p^2}$.
We compute the potential as being the difference between the 
Hamiltonian with and without a pair of static external charges separated by 
a distance $L$,
\bear
V(L)&=& H_q-H_0 =  -(L_q-L_0)\nonumber\\
&=& -q\int d^2x A_\mu \delta^{\mu 0}
\lbrack \delta (x^1+L/2)\delta (x^2) -\delta (x^1-L/2)\delta (x^2)
\rbrack \nonumber\\
&=& -q\lbrack A_0(x^1=-L/2,x^2=0)- A_0(x^1=L/2,x^2=0)\rbrack\; ,
\label{v-as-extsource}
\eear
where we have integrated over the two space components in order to
find the potential, and considered the source as corresponding to two
fixed charges of magnitude $q$ located at the points defined by the 
respective delta functions. Note that $L_q (L_0)$ denote the Lagrangeans in
the presence (absence) of the charges.

We now consider the equations of motion associated with the action
defined by means of (\ref{bosoaction}). 
The field equation in the covariant gauge reads
\be
-H(\partial=\sqrt{-\partial^2})  \epsilon _{\mu\nu\rho}\frac
{\partial^\nu}{\partial^2} A^\rho
+G(\partial)\left(
g_{\mu\nu}-\frac{\partial_{\mu} \partial_{\nu}}{\partial^2}\right) 
A^\nu +J_\mu =0 \label{a-eq-motion}
\ee

Defining the curl of $A_\mu$ as
\be 
{\cal A}_\mu = -\epsilon_{\mu\nu\beta}\partial^\nu A^\beta\label{curl}
\ee
the equation of motion can be expressed as
\be
\lbrack \Box_{nonloc} +m_{nonloc}^2\rbrack 
{\cal A}_\mu =-
\epsilon_{\mu\nu\beta}\partial^\beta J^\nu - 
f_{nonloc}J_\mu\label{eq-motion-curl-a}
\ee
where 
\bear
\Box_{nonloc}&=& G(\partial) \nonumber\\
m_{nonloc}^2&=& \frac{H^2 (\partial)}{\partial^2 G(\partial)}\\
f_{nonloc}&=&\frac {H(\partial)}{G(\partial)}\nonumber
\eear
In the absence of sources, and in the large $m$ limit\cite{abdban}, 
it reproduces the familiar massive mode of Maxwell-Chern-Simons
theory\cite{djt}. From (\ref{v-as-extsource}) 
it is seen that an expression for $A_0$ is required to calculate the 
potential. This is given in terms of the curl (\ref{curl}) 
by
\be {\cal A}_2=-\partial_1A_0\label{a2a0}
\ee

The time independent solution for ${\cal A}_2$ corresponding to the sources 
describing static quarks can be obtained from (\ref{eq-motion-curl-a}). 
Using this result with (\ref{a2a0}) 
finally yields, after integrating over the angular variables,
\be
A_0(t,L)=A_0(0,L) = - \frac{q^2}{\pi} \int^\infty_0 
\frac{k J_0 (kL)}{\tilde{G}(k)+\frac{\tilde{H}^2 (k)}{k^2
\tilde{G}(k)}} dk \label{sol-static-quarks}
\ee
The integration over the angular variables in the Fourier transformation
led to the Bessel function $J_0(kL)$; in the case where the denominator
is given by the familiar result (i.e. Feynman propagator, which also appears
in the large mass limit, see \cite{abdban}) the result of the integration
is just the modified Bessel function \cite{gradsh}.

The potential is now found from (\ref{v-as-extsource}), 
(\ref{eq-motion-curl-a}) 
and (\ref{sol-static-quarks}), 
reading
\be
V(L)= - \frac{q^2}{\pi} \int^\infty_0 
\frac{k J_0 (kL)}{\tilde{G}(k)+\frac{\tilde{H}^2 (k)}{k^2
\tilde{G}(k)}} dk \label{v-int}
\ee
where the functions $G(k)$ and $H(k)$ are given by the expressions
\bear
G(k)&=& k^2 +\frac {e^2 m}{4\pi}\left[ 1- \left(\frac {2m}k -\frac
  k{2m}\right)\,  \arctan \left( \frac
k{2m} \right) \right] \quad ;\label{func-g}\\
H(k)&=&\frac {e^2 m k}{2\pi} \, \arctan \left(\frac k{2m}\right)
\label{func-h}
\eear

The above equation takes a particularly simple form in the infinite
mass limit,
\be
V(L)=-{1\over \pi}{q^2\over 1+{ e^2\over 6\pi m}}K_0\left(\frac
  {e^2}{4\pi}L\right) 
\equiv -{q_{ren}^2\over \pi}K_0\left(\frac {e^2}{4\pi}L\right)
\label{v-as-bessel}
\ee
This reproduces our earlier results in \cite{abdban}.
The asymptotic form of the Bessel function signalizes screening.

For arbitrary mass however, a simple closed form expression cannot be 
obtained. We therefore have to use numerical methods. They are presented
as follows. We first plot the function  $V(L)$ given in (\ref{v-int}) 
as a function of $L$ for various values of the mass parameter $m$. The
result is plotted in figure 1. It is immediately obvious that
the screening effect is qualitatively independent of the mass, and the
quantitative dependence extremely small. Indeed, the graphs are almost bound
inside a rather narrow band defined by the results obtained for
$m=0$ and $m=\infty$.

\begin{figure}[h!]
\begin{center}
\leavevmode
\bearst
\epsfxsize= 12truecm{\epsfbox{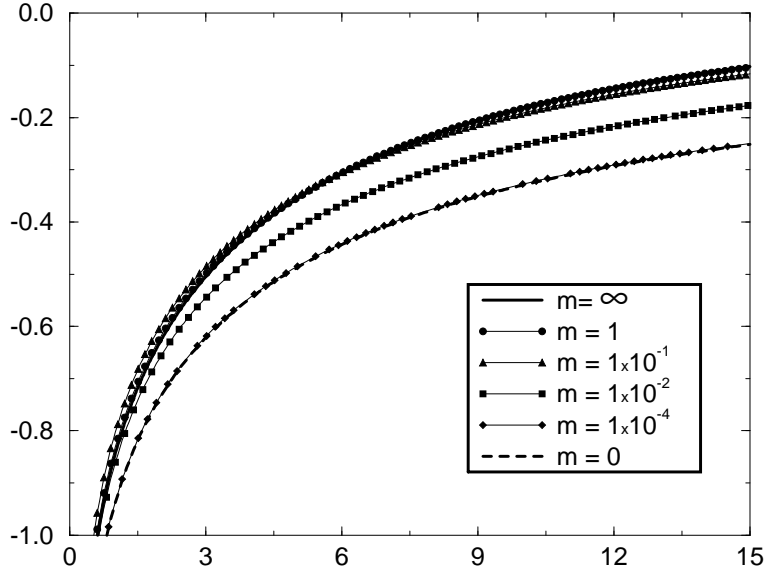}} 
\eearst 
\caption{{\small\it Potential as a function of the distance for
various values of the mass parameter. The value $m=1$ practically
coincides with the asymptotic value $m=\infty$. }}
\end{center}
\end{figure}

In general, due to the appearance of non algebraic functions the 
expressions appearing in (\ref{func-g}) and 
(\ref{func-h}) are rather clumsy. In reference \cite{fosco} 
simple expressions have been derived which, according to the authors, give 
a good approximation to these functions in the whole range of values of 
the parameter. The approximations are
\bear
G(k)&\approx & k^2\left\{ 1+\frac 1{16}\left[ k^2 +\left( \frac{3\pi m} 4
\right)^2\right]^{-\frac 12}\right\}\label{func-g-ap}\; , \\
H(k)&\approx &  \frac{e^2 m k^2}{4} \left(
k^2 + \pi^2 m^2
\right)^{-\frac 12}\label{func-h-ap}\quad . 
\eear

\begin{figure}[h!]
\begin{center}
\leavevmode
\bearst
\epsfxsize= 7truecm{\epsfbox{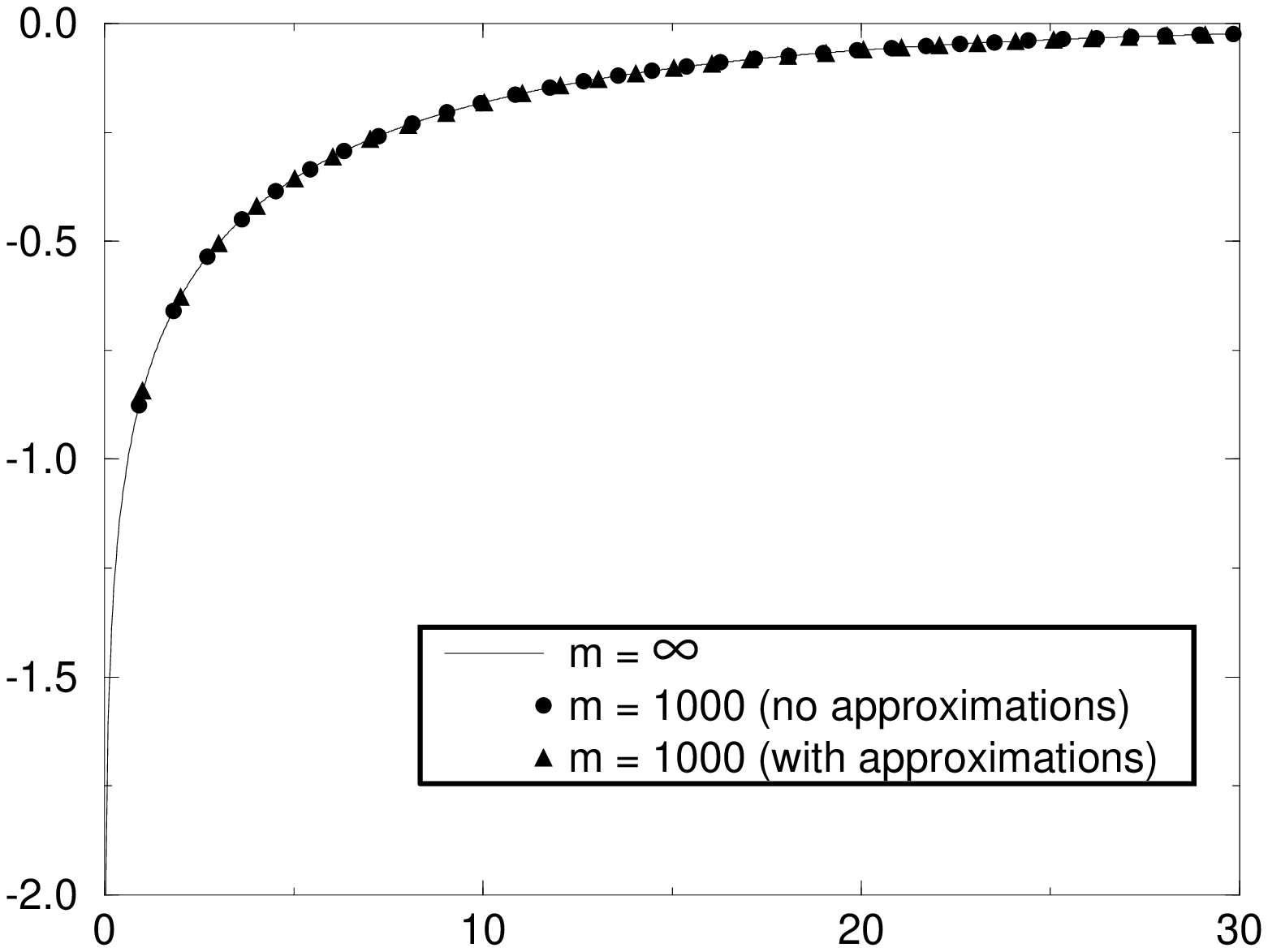}} \epsfxsize= 7truecm{\epsfbox{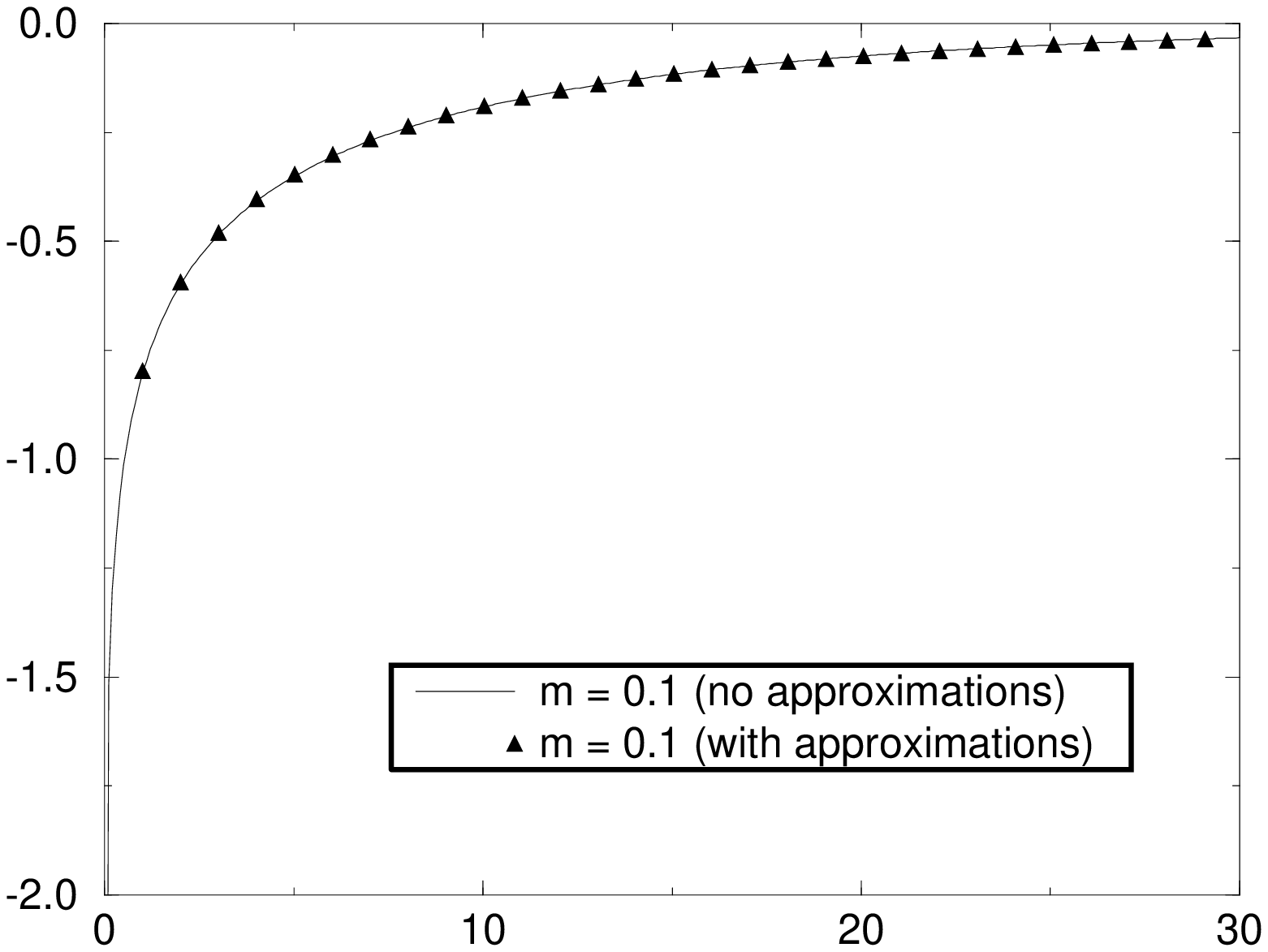}} 
\eearst 
\caption{{\small\it (Left) Potential function for infinite mass, and for $m=1
000$ using (\ref{func-g-ap}) and (\ref{func-h-ap}). (Right) Potential function for $m=0.1$ using the exact results
(\ref{func-g}) and (\ref{func-h}) and approximate forms
(\ref{func-g-ap}) and (\ref{func-h-ap}). }}
\label{fig:m=1000 & m=0.1}
\end{center}
\end{figure}

We tested this assumption for the computation of the potential, 
comparing the asymptotic result with the one obtained with the
approximations for $m=1000$. We repeated the procedure for $m=0.1$,
which shows similar findings.  The result is shown in figure
\ref{fig:m=1000 & m=0.1}
, indicating that the approximation agrees remarkably well with the
exact results. We also checked the approximations directly in figure
\ref{fig:GandH}
. Comparison of the expressions for $G$ e $H$ using
the approximations and the exact result shows that there is little
discrepancy. 

\begin{figure}[h!]
\begin{center}
\leavevmode
\bearst
\epsfxsize= 8truecm{\epsfbox{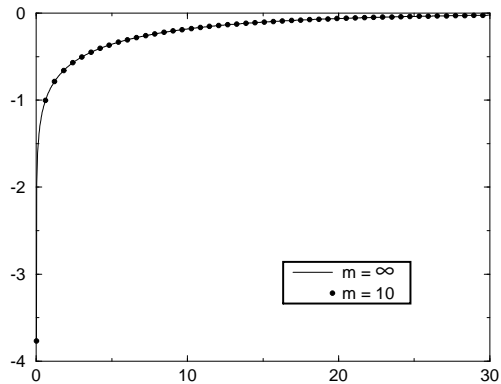}} 
\eearst 
\caption{{\small\it Comparison between the potential for $m=10$ and
the asymptotic result.}}
\label{fig:m=10}
\end{center}
\end{figure}

\begin{figure}[h!]
\begin{center}
\leavevmode
\bearst
\epsfxsize= 7truecm{\epsfbox{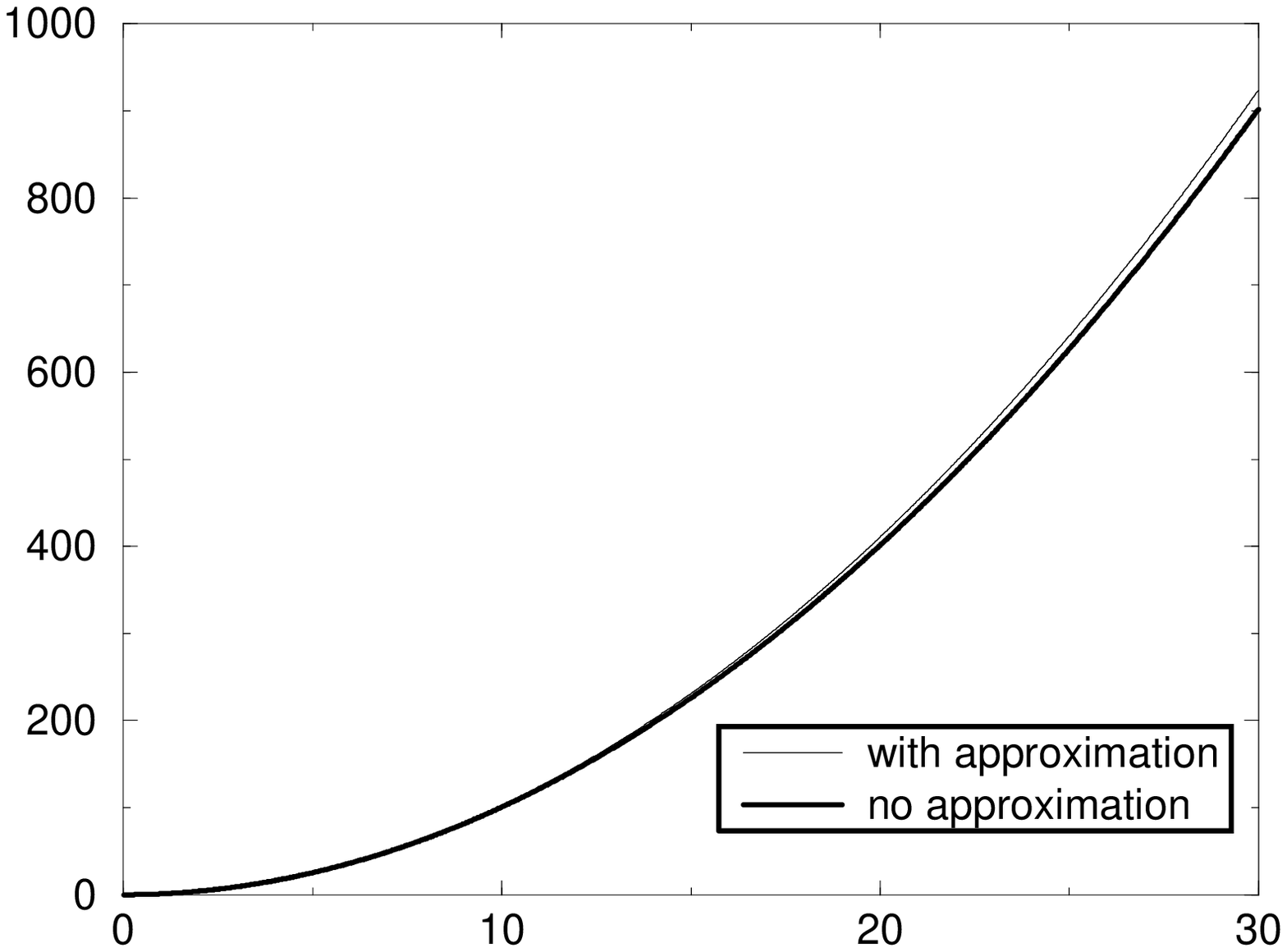}} \epsfxsize= 7truecm{\epsfbox{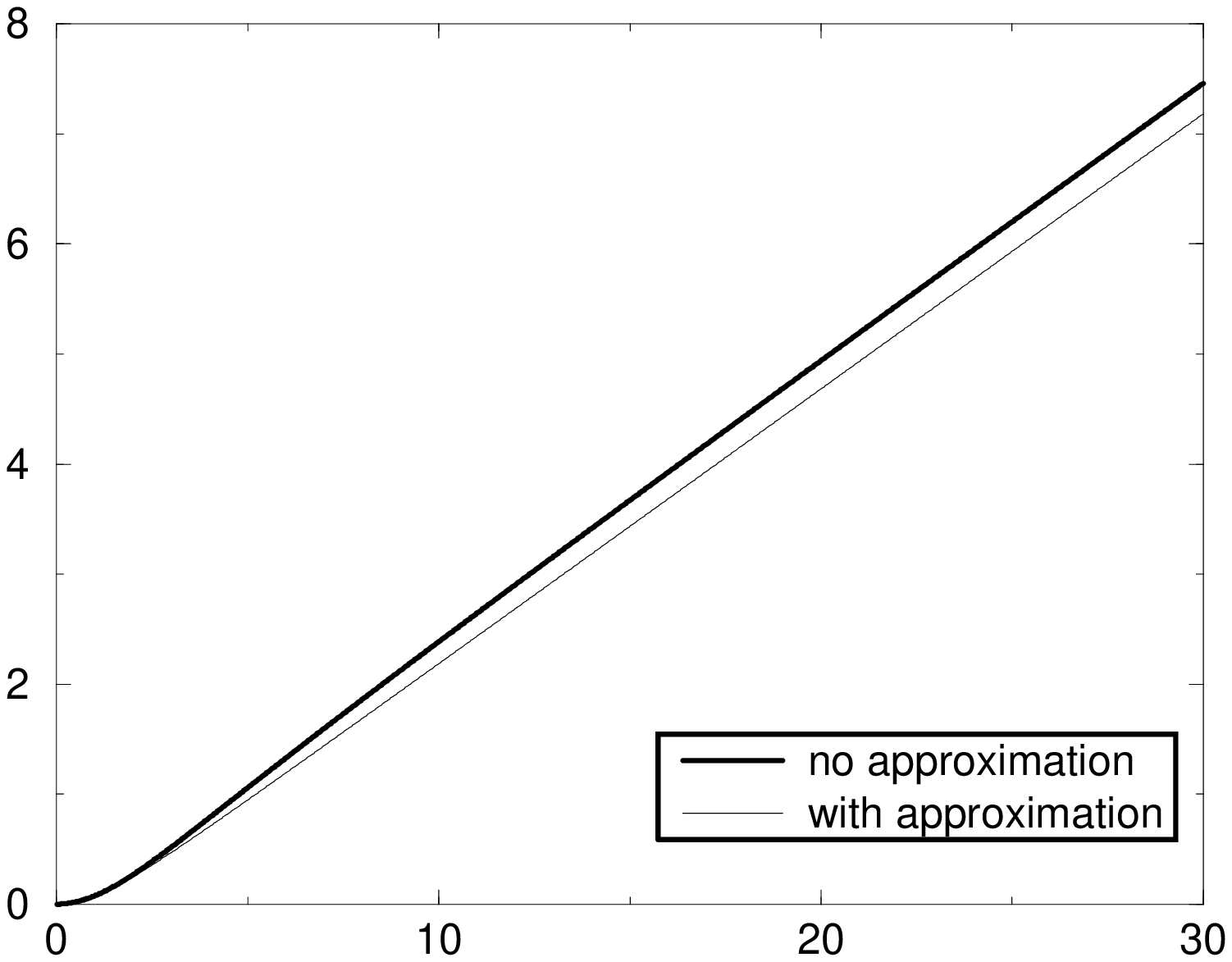}} 
\eearst 
\caption{{\small\it Function G (left) and H (right) with $m=1$, using the exact
definitions (\ref{func-g}) and (\ref{func-h}), and the approximations
(\ref{func-g-ap}) and (\ref{func-h-ap}).}} 
\label{fig:GandH}
\end{center}
\end{figure}

The approach to asymptotics in the computation of the potential has also
been analysed. We have verified that it is very quick. Indeed, for reasonably
low values of the mass the potential already shows the asymptotic value. We 
illustrated the behaviour in the case $m=10$ in figure \ref{fig:m=10}
, which was done with the exact expressions.

\section{Discussion}

Here we worked out an approach for obtainning the semiclassical
inter-quark potential for arbitrary values of the fermion mass parameter,
generalizing our previous work\cite{abdban}, where only the infinite mass 
approximation  was analysed. In that case, the expression
for the effective action is local, and the interquark potential
could be computed in closed form, showing explicitly that the
model lives in a screening phase. We attempted here to go beyond the
large mass limit. However, the expressions thus obtained are
nonlocal, and we had to resort to numerical simulations. The results are
nevertheless rewarding, especially in view of the extremely mild dependency
upon the fermion mass, {\it i.e.} almost every physical quantity related to
the screening potential is almost independent of the mass,
for $0\le m\le 1$ (we suppose $e=1$) and achieves the asymptotic value
already for $m$ of order unit.

The screening obtained for all values of the mass parameter supports the
observations obtained in two-dimensional QCD\cite{amz}, where the screening
phase also prevails almost universally (see also \cite{gross}).
This opens up the discussion for a large number of interesting possibilities.
In particular, it is interesting to stress that
the same mechanism may work for the non-abelian case in three dimensions,
since in the large mass limit the effective action turns out to be the
non abelian generalization of the Maxwell Chern Simons theory. In an
axial gauge, and in the weak coupling limit, the Maxwell Chern Simons
action coincides with the abelian counterpart, and the same conclusions
are expected. If one dares speculate that the mass dependence is as mild
as we have obtained in the above discussion, then all conclusions
can be carried over to the non abelian case as well, a tantalising result!
This would imply an almost universal screening behaviour for low 
dimensional systems. We hope to come back to these interesting questions
in a future work.

We also tested the approximative formulae (\ref{func-g-ap}) and 
(\ref{func-h-ap}), usually taken as
a good approximation within 10\% accuracy. The difference between  the
potential calculated using this approximations and the exact results
is less than $10^{-2}$, too small to be seen in figure
\ref{fig:m=1000 & m=0.1}. 

We finally comment on the infinite and zero mass limits of the model. 
In the latter case
we have to go beyond the quadratic approximation, since higher corrections
have to be computed in a gauge theory  where further
powers of the external momenta show up in the computation of diagrams.
One loop fermionic diagrams in gauge theories result
in powers of momentum and functions of momentum over mass, in general
$f(\frac {p^2}{m^2})$. While the limit $p\to 0$ and $m\to\infty$ is
unambiguous, the double limit $p\to 0$ and $m\to 0$ is not well defined,
and depends on the order they are taken. Therefore strong infrared
dependence on the mass may invalidate the procedure. Notice that the
discussion of screening pressuposes a large distance (namely small
momentum) limit, which may not commute with the zero mass limit. The
infinite mass limit obtained in the quadratic approximation is however
expected to survive even in the non-quadratic regime. We hope to come
back to these points in a future publication.

{\bf Acknowledgements}: this work  has been partially supported
by Conselho Nacional de Desenvolvimento Cient\'\i fico e Tecnol\'ogico,
CNPq, Brazil, and Funda\c c\~ao de Amparo \`a Pesquisa do Estado de
S\~ao Paulo (FAPESP), S\~ao Paulo, Brazil.


\end{document}